\def \be {\begin{equation}}
\def \ee {\end{equation}}
\def \bea {\begin{eqnarray}}
\def \eea {\end{eqnarray}}
\def \nn {\nonumber}

\def \la {\langle}
\def \ra {\rangle}
\def \R {{\bf R}}
\def \C {{\bf C}}
\def \Z {{\bf Z}}
\def \del {\partial}
\def \dels {\partial\kern-.5em / \kern.5em}
\def \As {{A\kern-.5em / \kern.5em}}
\def \Ds {D\kern-.7em / \kern.5em}

\def \a {\alpha}

\def \dag {\dagger}

\def \G {\Gamma}
\def \d {\delta}
\def \eps {\epsilon}

\def \k {\kappa}
\def \lam {\lambda}

\def \s {\sigma}

\def \th {\theta}

\def \II {I\hspace{-.1em}I\hspace{.1em}}

\def \IIB {\mbox{\II B\hspace{.2em}}}
\def \p {\partial}

\def \j {j}

\def \Nc {{\cal N}}

\def \Psit {\tilde{\Psi}}

\def \Dc {{\cal D}}

\documentclass[12pt]{article}

\begin{document}
\begin{titlepage}

\begin{center}
\hfill hep-th/0005268\\
\vskip .5in

\textbf{\large Large N Expansion From Fuzzy AdS$_2$ }

\vskip .5in
{\large Pei-Ming Ho$^1$, Miao Li$^{2,1}$}
\vskip 15pt

{\small \em $^1$ Department of Physics, National Taiwan
University, Taipei 106, Taiwan}

\vskip 15pt
{\small \em $^2$ Institute of Theoretical Physics, Academia Sinica,
Beijing 100080}

\vskip .2in
\sffamily{
pmho@phys.ntu.edu.tw\\
mli@phys.ntu.edu.tw}

\vspace{60pt}
\end{center}
\begin{abstract}
We study the quantum analogue of primary fields and their descendants
on fuzzy $AdS_2$, proposed in hep-th/0004072. Three-point vertices
are calculated and shown to exhibit the conventional $1/N$ expansion as
well as nonperturtive effects in large $N$,
thus providing a strong consistency check of the fuzzy $AdS_2$
model. A few new physical motivations for this model are also
presented.

\end{abstract}
\end{titlepage}
\setcounter{footnote}{0}

\section{Introduction}

Quantum mechanics and Einstein's general relativity are two major 
achievements in physics in twentieth century. In the last thirty years,
string theory has become the most promising theory unifying
these two otherwise highly irreconcilable theories. Nevertheless
in string theory, until quite recently, spacetime has been treated
as a continuous platform upon which one  builds up a perturbation
theory. It was suspected by many that when gravitational
interactions become strong, a continuous spacetime is doomed,
and one has to replace it by something else. In other words,
when nonperturbative effects are properly taken into 
account, spacetime in string theory will become fundamentally
fuzzy.

A concrete model in which space and time are quantized was proposed
very recently \cite{HL}. The classical counter-part is the
two dimensional anti-de Sitter space which one can obtain
as the near horizon limit of an extremal black hole in 4 
dimensions. Quantization of spacetime has a long history,
it was first proposed in \cite{SNYDER, YANG}. The motivation
at that time is to cure the UV divergences in QED, and
this is rendered unnecessary by the modern quantum field
theory.  However, very general arguments show that spacetime
are fundamentally noncommutative in string theory \cite{UNC}. And
that some unusual physics required to explain quantum
black holes also indicates that space and time are quantized
simultaneously \cite{THOOFT0}. One piece of such unusual physics
is teleology which was beautifully demonstrated in a recent
paper \cite{SST} in one of the simplest models of noncommutative
spacetime. Another manifestation of noncommutativity of
spacetime is the UV/IR relation in the AdS/CFT correspondence
\cite{UVIR}, indeed one of our motivations to propose the
fuzzy $AdS_2$ model is to derive holography from spacetime
noncommutativity. 

Note that quantum spheres in the AdS/CFT
correspondence were first proposed in \cite{JR, HRT} to
explain the stringy exclusion principle \cite{SEP}. A remarkable
physical mechanism for this phenomenon is proposed
in \cite{MST}, and subsequently it was shown in \cite{MIAO} that
this mechanism is compatible with the stringy spacetime
uncertainty relation, providing another indication of quantization 
of space and time. For even dimensional spheres, it
is shown in \cite{HL} that the quantum geometry of the sphere
is really a fuzzy sphere. In particular, for $S^4$, the
model already made its appearance in \cite{CLT, BV}.
The fuzzy $S^2$ is determined using the dipole mechanism and
the fuzzy $AdS_2$ is proposed based on analogy.

The fuzzy $AdS_2$ model resembles much the noncommutative horizon
of a 4 dimensional Schwarzschild black hole proposed 
by 't Hooft \cite{THOOFT0}. Its study and clarification
of many physics issues in this context will undoubtedly shed
much light on the quantum physics of black holes which have
been source of riddles as well as inspirations.

We shall make a first step in studying physics in the fuzzy
$AdS_2$ in this paper. The uncertainty relation results from
this model is unusual in that it is much larger than the
one showing up for instance in \cite{THOOFT0}. This is not surprising 
to us, since this model is a descendant of the 4 dimensional
extremal black hole whose quantum properties are unusual
too. The most convincing check that our model is correct is
the right pattern of $1/N$ expansion for the effective
interaction vertices. We will in the next section review some discussion
of \cite{ANDY} on the solution to the Laplace equation in
$AdS_2$, and in sect.3 review our definition of fuzzy $AdS_2$.
We fix the geometric data of the fuzzy $AdS_2$ in sect.4 and
solve the quantum analogue of the Laplace equation. Sect.5
computes the three-point vertices, and demonstrates $1/N$ expansion
in these vertices. Remarkably, there are also instanton effects such
as terms $\exp(-2\pi nN)$, similar to \cite{BG}.
In the
discussion section we  present a shock-wave argument for fuzzy
$AdS_2$.

We leave many interesting questions, such as how to understand 
holography, to future work.

\section{$AdS_2/CFT_1$ Duality}

In this section we review what has been discussed about
the $AdS_2/CFT_1$ duality \cite{ANDY}.

For a gravitational theory on $AdS_2$,
the isometry group $SL(2,\R)$ is enhanced to
half of the 1+1 dimensional conformal group.
We denote the Killing vectors for the isometry group
by $L_0, L_1, L_{-1}$.

For a scalar field $\Phi$ of mass $M$
propagating in the bulk of $AdS_2$,
its wave equation is
\be
\nabla^2\Phi=M^2\Phi.
\ee
The Laplacian $\nabla^2$ is just $4/R^2$ times
the Casimir of $SL(2,\R)$.
To get a primary solution of the wave equation,
we impose the constraints
\be
L_{1}\Phi_h=0, \quad L_0\Phi_h=h\Phi_h
\ee
on $\Phi$.
Then the solution is easily found to be
\be \label{sol}
\Phi_h=(e^{-2iu^+}-e^{-2iu^-})^h,
\ee
where we used the global coordinates
\be \label{global}
ds^2=-{4R^2\over\sin^2(u^+-u^-)}du^+du^-.
\ee
This solution is normalizable and has
\be
M^2=4h(h-1)/R^2.
\ee
The primary solution in the bulk corresponds to
a primary operator on the boundary CFT.

Other solutions of the wave equation with which
$\Phi_h$ forms an irreducible representation of $SL(2,\R)$
can be obtained by acting $L_{-1}$ on $\Phi_h$.
Denote the normalized solutions obtained this way by $\Psit_{hn}$,
where $n$ is the number of times $L_{-1}$ acts on $\Phi_h$.
This is a highest weight representation.
Its complex conjugation gives a lowest weight representation.
All together they form a complete basis with which
one can expand a scalar field of mass $M=2\sqrt{h(h-1)}/R$ as
\be
\Phi=\sum_{n=0}^{\infty}a_n^{\dag}\Psit_{hn}^{\dag}+a_n\Psit_{hn}.
\ee
The canonical commutation relation is
\be
[a_m, a_n^{\dag}]=\delta_{mn}.
\ee

As $h$ is roughly proportional to the mass of a scalar field,
it was shown that there should exist a cutoff of $h$ around the value
\be \label{cutoff}
h\sim 1/G_N,
\ee
where $G_N$ is the Newton constant in $AdS_2$.

\section{Fuzzy $AdS_2\times S^2$}

Consider the near horizon limit of a 4 dimensional charged
black hole in string theory \cite{MS}.
For instance, by wrapping two sets of membranes and
two sets of M5-branes in $T^7$, one obtains a 4D charged,
extremal black hole \cite{KT}.
The brane configuration is as follows.
Denote the coordinates of $T^7$ by $x_i$, $i=1, \dots, 7$.
A set of membranes are wrapped on $(x_1, x_2)$,
another set are wrapped on $(x_3, x_4)$.
A set of M5-branes are wrapped on $(x_1, x_3, x_5, x_6, x_7)$,
the second set are wrapped on $(x_2, x_4, x_5, x_6, x_7)$.
By setting all charges to be $N$, one finds the metric of
$AdS_2\times S^2$ for $(x_0,x_8,x_9,x_{10})$:
\bea \label{adsm}
&ds^2=l_p^2\left(-{r^2\over N^2}dt^2+{N^2\over r^2}dr^2
+N^2d\Omega_2\right),\\
&F=-N d\Omega_{1+1}
-N d\Omega_2,
\eea
where $l_p$ is the 4 dimensional Planck length,
$d\Omega_{1+1}$ and $d\Omega_2$ are the volume forms
on $AdS_2$ and $S^2$, respectively.
The field $F$ is just the linear combination of all
anti-symmetric tensor fields involved.
Note that here for simplicity,
we consider the most symmetric case in which
all the charges appearing in the harmonics $1+Q_il_p/r$
are just $N$ which in turn is equal to the number
of corresponding branes used to generate this potential.
As a consequence, the tension of the branes compensates 
the volume of the complementary torus.
This means that the size of each circle of $T^7$
is at the scale of the M theory Planck length.

The same space $AdS_2\times S^2$ can also be obtained by
taking the near horizon limit of the 4 dimensional
extremal Reissner-Nordtstr\"om solution.

\subsection{Fuzzy $S^2$}

In \cite{HL} we proposed that the $S^2$ part of
the $AdS_2\times S^2$ space is a fuzzy $S^2$ \cite{MADORE}
defined by
\be \label{S2}
[Y^a,Y^b]=i\epsilon^{abc}Y^c,
\ee
where $Y_a$'s are the Cartesian coordinates of $S^2$.
(We use the unit system in which $l_p=1$.)
This algebra respects the classical $SO(3)$ invariance.

The commutation relations (\ref{S2})
are the same as the $SU(2)$ Lie algebra.
For the $(2N+1)$ dimensional irreducible representation of $SU(2)$,
the spectrum of $Y_a$ is $\{-N, -(N-1), \cdots, (N-1), N\}$.
and its second Casimir is
\be
\sum_{a=1}^3 (Y_a)^2=N(N+1).
\ee
Since the radius of the $S^2$ is $N l_p$
(in the leading power of $N$),
we should realize the $Y_a$'s as $N\times N$ matrices
on this irreducible representation of $SU(2)$.

An evidence for this proposal is the following.
For a fractional membrane wrapped on $(x_1,x_3)$ or $(x_2,x_4)$,
it is charged under the $F$ field generated by a set of M5-branes.
Denote the polar and azimuthal angles of $S^2$ by $(\th,\phi)$.
The stable trajectories of the membrane
with angular momentum $M$ are discrete, and
\be
\cos\th=\frac{M}{N}.
\ee
Furthermore, they all have the same energy of $1/N$.
It follows that, since $M$ is conjugate to $\phi$,
$\cos\th$ and $\phi$ do not commute with each other
in the quantized theory.
The resulting Poisson structure on $S^2$ is precisely
that of the fuzzy sphere.

In \cite{THOOFT1}, it was proposed that in $2+1$ dimensions
the spacetime coordinates are quantized according to
\be \label{xyL}
[x,y]=\frac{i}{\cos^2\mu}L,
\ee
where $\cos\mu$ is related to the mass of the particle,
and $L$ is the angular momentum on the 2 dimensional space.
To complete the algebra we also have
\be
[L,x]=iy, \quad [L,y]=-ix.
\ee
This algebra is Lorentz invariant,
and its 3+1 dimensional generalization
was given by Yang \cite{YANG} much earlier.
Note that this algebra (\ref{xyL})
was proposed based on general grounds for
a gravitational theory in 2+1 dimensions,
and we are content with the fact that
it is actually a consequence of the fuzzy $S^2$ (\ref{S2})
for massless particles ($\cos\mu=1$),
where $Y_3$ acts on $Y_1$ and $Y_2$ as
the angular momentum operator.

\subsection{Fuzzy $AdS_2$}

In \cite{HL} we further proposed that the $AdS_2$ part
is also quantized.
Let $X^{-1}, X^0, X^1$ be the Cartesian coordinates of $AdS_2$.
The algebra of fuzzy $AdS_2$ is
\bea
&[X^{-1}, X^0]=-i X^1, \label{AdS21}\\
&[X^0, X^1]=i X^{-1}, \label{AdS22}\\
&[X^1, X^{-1}]=i X^0, \label{AdS23}
\eea
which is obtained from the fuzzy $S^2$ by a ``Wick rotation''
of the time directions $X^0,X^{-1}$.
The ``radius'' of $AdS_2$ is $R=N l_p$, so
\be \label{XXR}
\eta_{\mu\nu}X^{\mu}X^{\nu}=(X^{-1})^2+(X^0)^2-(X^1)^2=R^2,
\ee
where $\eta=$diag$(1,1,-1)$.
The isometry group $SL(2,\R)$ of the classical $AdS_2$
is a symmetry of this algebra,
and thus is also the isometry group of the fuzzy $AdS_2$.

For later use, we define the raising and lowering operators
\be \label{X+-}
X_{\pm}\equiv X^{-1}\pm i X^0,
\ee
which satisfy
\be
[X^1, X_{\pm}]=\pm X_{\pm}, \quad [X_+, X_-]=-2X^1,
\ee
according to (\ref{AdS21})-(\ref{AdS23}).

The radial coordinate $r$ and the boundary time coordinate $t$
are defined in terms of the $X$'s as
\be
r=X^{-1}+X^1, \quad t=\frac{R}{2}(r^{-1}X^0+X^0 r^{-1}),
\ee
where we symmetrized the products of $r^{-1}$ and $X^0$
so that $t$ is a Hermitian operator.
The metric in terms of these
coordinates assumes the form (\ref{adsm}).
It follows that the commutation relation for $r$ and $t$ is
\be \label{rt}
[r,t]=-iRl_p.
\ee

The following simple heuristic argument also
suggests this commutation relation.
Consider a closed string in $AdS_2$.
(Since the space is one dimensional,
the closed string actually looks like an open string
with twice the tension.)
Take the Nambu-Goto action for a fractional string of tension $1/N$
and take the static gauge $t=p_0\tau$.
It follows that the action is
\bea
S&=&\frac{1}{2\pi N\a'}\int_{-\infty}^{\infty}dt\int_0^{2\pi}d\s
    \sqrt{(p_0\dot{r}^2)} \nn\\
 &=&\frac{1}{\pi N\a'}\int dt\int_0^{\pi} p_0|\dot{r}| \nn\\
 &=&\frac{1}{N\a'}\int d\tau \dot{r}\frac{\del t}{\del\tau},
\eea
where we have assumed that $\dot{r}>0$ for $0<\s<\pi$
and $\dot{r}<0$ for $\pi<\s<2\pi$.
The last line above shows clearly that $r$
is the conjugate variable to $t$ up to a factor of $N\a'=R l_p$,
and so the quantization of this fundamental string is (\ref{rt}).
We will set $l_p=1$ in the rest of the paper.

\section{Properties of Fuzzy $AdS_2$}

One can realize the algebra (\ref{AdS21})-(\ref{AdS23}),
which is the same as the Lie algebra of $SL(2,\R)$,
on a unitary irreducible representation of $SL(2,\R)$.
The question is which representation is the correct choice.
As we show in appendix \ref{reps},
since the range of $X_1$ goes from $-\infty$ to $\infty$ for $AdS_2$;
when $R>1/2$, the proper choice is the principal continuous series,
and when $R<1/2$, it is the complementary series.
Since we have $R=N>1$ for our physical system,
we should consider the principal continuous series only.
A representation in this series is labeled by two parameters
$\j=1/2+is$ and $\a$, where $s, \a$ are real numbers,
and $0\leq\a<1$.
The label $\j$ determines the second Casimir as
\be
c_2=\eta_{\mu\nu}X^{\mu}X^{\nu}.
\ee
It follows from (\ref{XXR}) and $R=N$ that one should take
\be \label{jN}
\j=1/2+iN.
\ee
However, strictly speaking, we can only be sure of these relations
in the leading power of $N$.

What is the physical interpretation
of the other label $\a$ of these representations?
We will argue in sect.\ref{interaction} that
$\a$ corresponds to the VEV of axion
when viewing this system as \IIB string theory via duality.
However, for most of the discussions below,
we will focus on the representation with $(\j=1/2+iN, \a=0)$,
to be denoted by $\Dc_N$.
This is the case for which the reflection symmetry
$X^1\rightarrow -X^1$ is not broken.

Functions on the fuzzy $AdS_2$ are functions of the $X$'s.
They form representations of the isometry group $SL(2,\R)$.
The three generators $L_{\mu\nu}$ of the isometry group act on $X$ as
\be
[L_{\mu\nu}, X^{\k}]=i(\d_{\nu}^{\k}X_{\mu}-\d_{\mu}^{\k}X_{\nu}),
\ee
where $X_{\mu}=\eta_{\mu\nu}X^{\nu}$.
A very interesting property of the algebra of fuzzy $AdS_2$ is that
the action of the generators $L_{\mu\nu}$ is
the same as the adjoint action of the $X_{\k}$.
That is,
\be
[L_{\mu\nu}, f(X)]=\eps_{\mu\nu\k}[X^{\k}, f(X)]
\ee
for an arbitrary function $f(X)$.
The operators $L_{\mu\nu}$ act on the functions as differential operators.

The integration over the fuzzy $AdS_2$ is just the trace
over the representation $\Dc_N$
\be
\int f(X)\equiv c\mbox{Tr}(f(X))=c\sum_{n\in\Z}\la n|f(X)|n\ra,
\ee
where $c$ is a real number.
This integration is invariant under $SL(2,\R)$ transformations.
In the large $N$ limit, $c_2\gg 1$,
the trace can be calculated and its comparison with
an ordinary integration on the classical $AdS_2$
with metric (\ref{adsm}) shows that
\be
c=2\pi N
\ee
in the leading power of $N$.
The inner product of two functions, as well as the norm of a function
are defined by integration over the fuzzy $AdS_2$ in the usual way:
\be
\la f(X)|g(X)\ra=\int f^{\dag}(X)g(X),
\quad \|f(X)\|^2=\int f(X)^{\dag}f(X).
\ee

In view of organizing the functions into representations of $SL(2,\R)$,
in order to describe the boundary CFT dual to the bulk theory
on fuzzy $AdS_2$ via holography,
we derived all functions corresponding to the lowest
and highest weight states in the principal discrete series.
The information about the underlying fuzzy $AdS_2$,
i.e., the value of $N$,
is encoded in the precise expressions of these functions.

Denote the lowest weight state by $\Psi_{\j\j}$, or just $\Psi_\j$.
The states of higher weights $\Psi_{\j m}$ ($m>\j$)
in the same irreducible representation are obtained as
\be \label{X+X+}
[X_+,[X_+,\cdots[X_+,\Psi_\j]\cdots]],
\ee
where $X_+$ appears $(m-\j)$ times.
In appendix \ref{functions}, we find the explicit expressions
for the functions $\Psi_\j$ as
\be
\Psi_\j=\left(\frac{1}{X^1(X^1-1)+c_2}X_+\right)^\j.
\ee
In the large $N$ limit, using $c_2=R^2$ and
the following coordinate transformation
\be
X^1=R\cot(u^+ - u^-),
\quad X_{\pm}=\frac{R}{\sin(u^+ - u^-)}e^{\mp i(u^+ + u^-)},
\ee
one finds
\be
\Psi_\j\rightarrow \left(\frac{e^{-2iu^+}-e^{-2iu^-}}{-2iR}\right)^\j,
\ee
where $u^+$, $u^-$ are the coordinates appearing in (\ref{global}),
in agreement with (\ref{sol}).

Let
\be \label{Im-def}
I_{\j m}\equiv \mbox{Tr}(\Psi^{\dag}_{\j m}\Psi_{\j m}),
\ee
then the normalized states are
\be
\Psit_{\j m}\equiv \frac{1}{\sqrt{cI_{\j m}}}\Psi_{\j m}.
\ee
As a normalized basis of an $SL(2,\R)$ representation,
they satisfy
\bea
{[X_+,\Psit_{\j m}]}&=&a_{\j(m+1)}\Psit_{\j(m+1)}, \\
{[X_-,\Psit_{\j m}]}&=&a_{\j m}\Psit_{\j(m-1)},
\eea
where
\be
a_{\j m}=\sqrt{m(m-1)-\j(\j-1)}.
\ee
The explicit expressions of $I_m\equiv I_{mm}$ are given
in appendix \ref{Im}.
They will be used later when we try to
extract physical information from the fuzziness of the $AdS_2$.

For a field $\Phi$ in the bulk of the fuzzy $AdS_2$,
one can decompose it into the basis functions $\Psit_{\j m}$ as
\be \label{jm}
\Phi(X)=\sum_{\j m}\phi_{\j m}\Psit_{\j m}(X),
\ee
where $\phi_{\j m}$ are the creation/annihilation operators
of the physical state with the wave function $\Psit(X)$.
By holography, these states are identified with those
in the boundary theory, which is a one-dimensional theory.

An interesting question for a wave function on
noncommutative theory is how to define expectation values.
For instance, should the expectation value of $X_1$ for
a wave function $\Psi$ be
\be \label{X1}
\mbox{(1)}\;\int \Psi^{\dag}X^1\Psi,
\quad \mbox{(2)}\int \Psi X^1\Psi^{\dag},
\quad \mbox{or}\;\; \mbox{(3)}
\frac{1}{2}\int(X^1\Psi^{\dag}\Psi+\Psi\Psi^{\dag}X^1)\mbox{?}
\ee
The answer is that it depends on how one measures it.
If one measures the $X^1$ location of the wave function
according to its interaction with another field $\Phi$
under control in the experiment,
and if the interaction is described in the action by a term like
\be
\int \Psi^{\dag}\Phi\Psi,
\ee
then we expect that the choice (1) is correct.
But if the interaction is written differently,
the definition of expectation value should be modified accordingly.

Fixing a definition of the expectation value of $X_1$,
such as case (1) in (\ref{X1}),
one finds that for the same $\j$, the larger $m$ is,
which means larger energy at the boundary,
the larger is the expectation value of $X_1$
for the wave function $\Psi_{\j m}$.
This can also be viewed as a manifestation of
the UV/IR relation in $AdS$ space.

\section{Interaction in Fuzzy $AdS_2$} \label{interaction}

To see how the noncommutativity of the fuzzy $AdS_2$
incorporate physical data, presumably including the effect of
string quantization on $AdS_2$,
we consider an interaction term in the action of the bulk theory
of the form
\be
S_I=\lam\int\Phi^{\dag}_1\Phi_2\Phi_3,
\ee
where $\lam$ is the coupling constant for this three point interaction.

Expanding all three $\Phi_i$'s as (\ref{jm}) in the action,
one obtains the vertex
\be \label{vertex}
c\lam\mbox{Tr}(\Psit^{\dag}_{\j_1 m_1}\Psit_{\j_2 m_2}\Psit_{\j_3 m_3})
\ee
for the three states $(\phi_1)_{\j_1 m_1}$, $(\phi_2)_{\j_2 m_2}$
and $(\phi_3)_{\j_3 m_3}$.
Obviously, due to the isometry, the vertex vanishes unless
$m_1=m_2+m_3$.

For simplicity, consider the special case where all three
states participating the interaction are lowest weight states.
Then the vertex (\ref{vertex}) in question is $\lam V_{m_1 m_2 m_3}$,
where (for $m_1=m_2+m_3$)
\be \label{V2}
V^2_{m_1 m_2 m_3}=\frac{1}{c}\frac{I_{m_1}}{I_{m_2}I_{m_3}}.
\ee
In appendix \ref{Im}, we find that
\be
I_m=\frac{(2m-2)!}{((m-1)!)^2}
\left[\Pi_{k=1}^{m-1}\frac{1}{k^2-1+4c_2}\right]I_1,
\ee
where
\be
I_1=\frac{\pi}{\sqrt{c_2-1/4}}\mbox{tanh}\left(\pi\sqrt{c_2-1/4}\right).
\ee

We therefore have the large N expansion of the vertex (\ref{V2}).
Using (\ref{jN}), one finds (for $m_1=m_2+m_3$)
\be
V^2_{m_1 m_2 m_3}=\Nc_{m_2 m_3}
\frac{\left[\Pi_{\j_2=1}^{m_2-1}(1+\j_2^2/4N^2)\right]
\left[\Pi_{\j_3=1}^{m_3-1}(1+\j_3^2/4N^2)\right]}
{8\pi^2 N^2\left[\Pi_{\j_1=1}^{m_1-1}(1+\j_1^2/4N^2)\right]}
\left(1+2\sum_{n=1}^{\infty}e^{-2\pi nN}\right),
\ee
where
\be
\Nc_{m_2 m_3}=\frac{[2(m_1-1)]! [(m_2-1)!]^2 [(m_3-1)!]^2}
{[(m_1-1)!]^2 [2(m_2-1)]! [2(m_3-1)]!}.
\ee
With the possibility of corrections to (\ref{jN}) of order $(1/N)^0\sim 1$,
the $1/N$ expansion of the vertex is of the form
\be \label{Vm}
V_{m_1 m_2 m_3}\sim \frac{K}{N}\left(1+
\sum_{n=1}^{\infty}\frac{a_n}{N^{2n}}+
\sum_{n=1}^{\infty}e^{-2\pi nN}\sum_{k=0}^{\infty}
\frac{b_{nk}}{N^{2k}}\right).
\ee

This expression is reminiscent of a correlation function
in the case of type \IIB strings on $AdS_5\times S^5$ \cite{BG}.
It calls for an analogous interpretation.

The $1/N^2$ expansion in (\ref{Vm}) suggests that
the coupling constant in $AdS_2$ is of order $1/N^2$.
This is indeed the case.
The 11 dimensional Newton constant is just $1$ in Planck units.
Compactifying on $S^2\times T^7$ of size $4\pi N^2$
results in a dimensionless effective Newton constant
of order $1/N^2$ in $AdS_2$.

The overall factor of $1/N$ in (\ref{Vm})
is what one expects for a three-point correlation function.
The Newton constant is $1/N^2$ in $AdS_2$ as we just mentioned,
on the other hand it is also $g_s^2$ from
the viewpoint of a string theory on $AdS_2$.
This means that $g_s\sim 1/N$.
While the coupling constant $\lam$ in the bulk theory
should not depend on $N$ since it should not depend on
the geometry of the space it lives on,
the vertex (\ref{vertex}) should have the leading dependence on $N$
of a three-point function for closed string states,
which is of order $g_s\sim 1/N$.
This is exactly the dependence we find for the vertex (\ref{Vm}).

Finally, we identify the terms in (\ref{Vm}) proportional to
$\exp(-2\pi nN)$ as contributions from instantons.
This implies that the action of a single instanton equals $2\pi N$.
We have just argued that the string coupling constant $g_s$
is of order $1/N$.
Since the D-instanton action is $2\pi/g_s$,
it is precisely $2\pi N$ as we wish.

Similarly, we can consider $n$-point interaction vertex
in the bulk theory on $AdS_2$:
\be
S_n=\lam\int\Phi^{\dag}_1\Phi_2\cdots\Phi_n.
\ee
The leading dependence of the vertex will be $1/N^{n-2}$,
which is exactly what it should be for an $n$-point function
in string theory with coupling constant $g_s\sim 1/N$.

We conjecture that for M theory compactified on $AdS_2\times S^2$,
the perturbative as well as nonperturbative effects
of string quantization are encoded
in the noncommutativity of the fuzzy $AdS_2\times S^2$,
in the sense that the low energy effective theory
is most economically written as a field theory on
this noncommutative space.

Finally we make a comment on the other label $\a$
for a representation in the principal continuous series.
We have set $\a=0$ in the above.
When $\a$ is turned on, the expression of $a_{\j(m+\a)}$ becomes
\be
a_{\j(m+\a)}=\sqrt{(m-1/2+\tau)(m-1/2+\bar{\tau})}
\ee
for $m\in\Z$, where
\be
\tau=\a+iN\simeq\a+\frac{i}{g_s}.
\ee
The appearance of this combination of $\a$ with $g_s$
suggests the identification of $\a$ with $\chi/2\pi$,
where $\chi$ is the VEV of the axion field in \IIB theory.
While $\chi$ is only defined up to $2\pi$,
$\a$ is also defined only up to an integer.

\section{Discussion}

Although we already argued for the spacetime uncertainty
in the fuzzy $AdS_2$ from the general stringy uncertainty,
it should be interesting to compare the way 't Hooft
introduces spacetime noncommutativity in his S-matrix
ansatz \cite{THOOFT0, THOOFT2}. The basic method is
to study the effect of a shock-wave on a particle moving
in the opposite direction.

Use the global coordinates (\ref{global}). The metric
induced by a left-moving shock-wave assumes the form
\be \label{shock}
ds^2=-e^{2\phi}du^+du^-+h(du^+)^2,
\ee
where 
$$e^{2\phi}={4R^2\over \sin^2(u^+-u^-)}.$$
The scalar curvature is perturbed by a term
\be
{1\over 2}e^{-2\phi}\p_-(e^{-2\phi}\p_-h),
\ee
and the Einstein equation with a constant negative 
curvature is solved provided
\be
h=\cot(u^+-u^-)g(u^+)+f(u^+).
\ee
In the full 4 dimensions, we expect another Einstein
equation of the form $G_{++}=8\pi G_4T_{++}$, where
$G_4$ is the 4D Newton constant. Now $G_{++}\sim
h{\cal R}$, ${\cal R}$ is the scalar curvature. For a 
stress tensor $T_{++}$ proportional
to $\delta (u^+-x^+)$, the only solution is $g(u^+)=0$
and
\be
h\sim l_p^2\delta (u^+-x^+).
\ee
The proportionality constant is determined by how the
stress tensor is normalized. For a S-wave shock-wave
smeared over $S^2$, it is  $p_+/R^2$. However,
the dipole mechanism of \cite{HL} seems to indicate
that a shock-wave must be localized on a strip
on $S^2$ whose area is proportional to $Nl_p^2$. If true,
we expect
\be
\int du^+ T_{++}\sim {p_+\over Rl_p},
\ee
and this leads to
\be
h\sim Rl_pp_+\delta (u^+-x^+).
\ee
Now the shift on $u^-$ induced on a right-moving particle 
by the shock-wave is
\be
\Delta u^-\sim {p_+\over N}\sin^2(x^+-u^-)
\ee
as can be computed using (\ref{shock}). This shift
suggests a commutator
\be
[u^+,u^-]\sim {i\over N}\sin^2(u^+-u^-)
\ee
the one that is compatible with our fuzzy $AdS_2$ model.

\section*{Acknowledgment}

This work is supported in part by
the National Science Council, Taiwan, 
and the Center for Theoretical Physics at National Taiwan University.
The work of M.L. is also
supported by a ``Hundred People Project'' grant.

\appendix

\section{Representations of $SL(2,\R)$} \label{reps}

The algebra of fuzzy $AdS_2$ can be realized on an
irreducible representation of $SL(2,\R)$ ($SU(1,1)$).
Similar to the case of $SU(2)$,
a representation of $SL(2,\R)$ can be easily constructed
using the raising and lowering operators (\ref{X+-}).
The $X$'s act on the basis $\{|m\ra\}$ as
\bea
X^1|m\ra&=&m|m\ra, \\
X_+|m\ra&=&a_{m+1}|m+1\ra, \\
X_-|m\ra&=&a_m|m-1\ra,
\eea
where
\be
a_m=\sqrt{m(m-1)+c_2}.
\ee
The second Casimir is
\be \label{Casimir}
c_2=(X^{-1})^2+(X^0)^2-(X_1)^2=-\j(\j-1),
\ee
which should be identified with $R^2=N^2$.

However, not all such representations are unitary representations.
They are unitary only in the following four situations.
\begin{enumerate}
\item Principal discrete representations (lowest weight):\\
$\j\in\R$, $\j>0$, $\a=0$ and $m=\j, \j+1, \j+2, \cdots$.
\item Principal discrete representations (highest weight):\\
$\j\in\R$, $\j>0$, $\a=0$ and $m=-\j, -\j-1, -\j-2, \cdots$.
\item Principal continuous representations:\\
$\j=1/2+is$, where $s\in\R$, and without loss of generality,
one can choose $\a$ such that $0\leq\a<1$.
The eigenvalues of $X^1$ are $m=\a, \a\pm 1, \a\pm 2, \cdots$.
\item Complementary representations:\\
$1/2<\j<1$ and $(\j-1/2)<|\a-1/2|$ ($\j\in\R$),
where again we assumed that $0\leq\a<1$.
The eigenvalues of $X^1$ are again $m=\a, \a\pm 1, \a\pm 2, \cdots$.
\end{enumerate}

All unitary irreducible representations of $SL(2,\R)$
fit in these four cases, except the trivial identity representation.
The two principal discrete representations correspond
to a fuzzy version of the de Sitter space,
which can be obtained as one of the two hypersurfaces
in $2+1$ dimensional Minkowski space by requiring that
$(X^{-1})^2+(X^0)^2-(X^1)^2=-R^2$.
This can be seen by examining the spectrum of $X^1$.
On the other hand, the spectrum of $X^1$ for
the other two series are both extending from $-\infty$ to $\infty$.
Furthermore, from (\ref{Casimir}) we see that
For $R>1/2$ we have to choose the principal continuous series,
and for $R<1/2$ the complementary series.

\section{Functions on Fuzzy $AdS_2$
as Rep.s of $SL(2,\R)$} \label{functions}

A function $\Psi$ on the fuzzy $AdS_2$ corresponds to a lowest weight state if
\be \label{lowest}
[X_-, \Psi]|n\ra=0 \quad \mbox{for all}\;\; n\in\Z.
\ee
Take the ansatz for the lowest weight state in a principal discrete
representation with $\j=m\in\Z$ as
\be \label{Psi}
\Psi_m=f_m(X^1)X_+^m.
\ee
One can solve the function $f_m(X^1)$ up to an overall normalization
by imposing (\ref{lowest}).
The result is
\bea
f_m(X^1)
        &=&\Pi_{k=1}^m\frac{1}{(X^1-k+1)(X^1-k)+c_2}
           \label{fm} \nn \\
        &=&\frac{\G(X^1-m+(1-\s)/2)\G(X^1-m+(1+\s)/2)}
                {\G(X^1+(1-\s)/2)\G(X^1+(1+\s)2)},
\eea
where $\s=\sqrt{1-4c_2}$.
Inserting this back to (\ref{Psi}),
we find
\be \label{Psi0}
\Psi_m=\left(\frac{1}{X^1(X^1-1)+c_2}X_+\right)^m.
\ee

States in the same representation of higher weights
can be obtained by taking commutators of $X_+$ with $\Psi_m$ as
in (\ref{X+X+}).

The highest weight representations in the principal discrete series
can be obtained in a similar way, or by simply taking
Hermitian conjugation of the functions in the lowest weight representations.

\section{Calculation of $I_m$} \label{Im}

{}From (\ref{Psi0}) and (\ref{Im-def}), we find
\bea
I_m(x)&=&\sum_{n=-\infty}^{\infty}\la n|\Psi_m^{\dag}\Psi_m |n\ra \nn\\
      &=&\sum_{n=-\infty}^{\infty}\Pi_{k=1}^m\frac{1}{a_{m+k}^2(x)},
\eea
where
\be
a_m^2(x)=m(m-1)+x.
\ee
In the expressions above we view $I_m$ as a function of
the Casimir $c_2$ and we have denoted $c_2$ by $x$.

The expression of $I_m(x)$ can be reduced algebraically to
an expression involving only $I_1$ times a function of $x$.
While generic $I_m$ can be calculated similarly,
we show here only the case of $I_2$:
\bea
I_2(x)&=&\sum_{n=-\infty}^{\infty}\frac{1}{(n(n-1)+x)(n(n+1)+x)} \nn\\
      &=&\sum_n\frac{1}{2n}(\frac{1}{n(n-1)+x}-\frac{1}{n(n+1)+x}) \nn\\
      &=&\sum_n\frac{1}{2}(\frac{1}{n}-\frac{1}{n-1})\frac{1}{n(n-1)+x} \nn\\
      &=&\sum_n\frac{-1}{2}\frac{1}{n(n-1)(n(n-1)+x)} \nn\\
      &=&\sum_n\frac{-1}{2x}(\frac{1}{n(n-1)}-\frac{1}{n(n-1)+x}) \nn\\
      &=&\frac{-1}{2x}\left(\sum_n(\frac{1}{n-1}-\frac{1}{n})
                           -\sum_n\frac{1}{n(n-1)+x}\right) \nn\\
      &=&\frac{1}{2x}I_1.
\eea
This derivation is not rigorous because terms involving $1/n$
appear and it diverges at $n=0$.
However it suffices to show the idea.
It is easy to modify it to get a well-defined, rigorous derivation
which leads to exactly the same final result.
The general expression for $I_m$ is
\be \label{Im0}
I_m(x)=\frac{(2m-2)!}{[(m-1)!]^2}
\left[\Pi_{k=1}^{m-1}\frac{1}{k^2-1+4c_2}\right]I_1(x).
\ee
It is just $I_1$ divided by a polynomial of $x$ of order $(m-1)$.
So our task is reduced to calculate $I_1(x)$.

Viewing $I_1(x)$ as a holomorphic function of $x\in\C$,
one can convince him/herself of the following equality
\be \label{I1}
I_1(x)=\frac{\pi}{\sqrt{x-1/4}}\mbox{tanh}\left(\pi\sqrt{x-1/4}\right)
\ee
by checking that the two sides of the equal sign
share the following properties:\\
(1) All the poles they have are at $x=-n(n-1)$ with the residue $2$
for all integers $n\in\Z$.\\
(2) They have zeros at $x=-n(n-1)$ for any half-integer $n$ except $1/2$.\\
(3) They equal $\pi^2$ at $x=1/4$.\\
(4) They approach the function $\pi/\sqrt{x}$ as $x\rightarrow\infty$.

\eject

\end{document}